\documentclass[aps,reprint,footinbib,amsmath,amssymb,amsfonts,superscriptaddress]{revtex4-1}

\usepackage{graphicx}
\usepackage[hypertexnames=false]{hyperref}

\newcommand{\mc}[1]{\ensuremath{\mathcal{#1}}}
\newcommand{\mr}[1]{\ensuremath{\mathrm{#1}}}

\newcommand{\tr}{\ensuremath{\mathrm{Tr}}}

\newcommand{\la}{\ensuremath{\langle}}
\newcommand{\ra}{\ensuremath{\rangle}}

\newcommand{\bra}[1]{\ensuremath{\langle #1 |}}
\newcommand{\ket}[1]{\ensuremath{| #1 \rangle}}
\newcommand{\prj}[1]{\ensuremath{| #1 \rangle\!\langle #1 |}}
\newcommand{\ovl}[2]{\ensuremath{\langle #1 | #2 \rangle}}
\newcommand{\epv}[2]{\ensuremath{\langle #1 | #2 | #1 \rangle}}
\newcommand{\matel}[3]{\ensuremath{\langle #1 | #2 | #3 \rangle}}

\newcommand{\ua}{\ensuremath{{\uparrow}}}
\newcommand{\da}{\ensuremath{{\downarrow}}}

\newcommand{\Eq}[1]{Eq.~\eqref{#1}}

\newcommand{\Fig}[1]{Fig.~\ref{#1}}

\newcommand{\Sec}[1]{Sec.~\ref{#1}}
\newcommand{\App}[1]{App.~\ref{#1}}

\usepackage{multirow}

\usepackage[normalem]{ulem} 
\usepackage{color} 

\allowdisplaybreaks 

\begin{document}
\title{Numerical renormalization group method for entanglement negativity at finite temperature}
\author{Jeongmin Shim}
\affiliation{Department of Physics, Korea Advanced Institute of Science and Technology, Daejeon 34141, Korea}
\author{H.-S. Sim}\email[]{hssim@kaist.ac.kr}
\affiliation{Department of Physics, Korea Advanced Institute of Science and Technology, Daejeon 34141, Korea}
\author{Seung-Sup B. Lee}\email[]{s.lee@lmu.de}
\affiliation{Physics Department, Arnold Sommerfeld Center for Theoretical Physics, and Center for NanoScience,
Ludwig-Maximilians-Universit\"{a}t, Theresienstra{\ss}e 37, D-80333 M\"{u}nchen, Germany}
\date{\today}

\begin{abstract}
We develop a numerical method to compute the negativity,
an entanglement measure for mixed states,
between the impurity and the bath in quantum impurity systems at finite temperature.
We construct a thermal density matrix by using the numerical renormalization group (NRG),
and evaluate the negativity by implementing the NRG approximation that reduces computational cost exponentially.
We apply the method to the single-impurity Kondo model and the single-impurity Anderson model.
In the Kondo model, the negativity exhibits a power-law scaling at temperature much lower than the Kondo temperature and a sudden death at high temperature.
In the Anderson model, the charge fluctuation of the impurity contribute to the negativity even at zero temperature when the on-site Coulomb repulsion of the impurity is finite,
while at low temperature the negativity between the impurity spin and the bath exhibits the same power-law scaling behavior as in the Kondo model.
\end{abstract}
\maketitle

\section{Introduction}\label{100}

Entanglement is a truly non-classical correlation~\cite{Plenio07,Guhne09,Horodecki09}, 
which often appears in many-body systems at macroscopic scale~\cite{Amico08,Eisert10,Laflorencie16}.
It can be quantified by various entanglement measures~\cite{Plenio07,Guhne09,Horodecki09},
and useful to understand many-body phenomena such as topological order~\cite{Kitaev06,Levin06} and quantum criticality~\cite{Vidal03}. 
The Kondo effect,
a many-body pheonomenon in quantum impurity systems induced by the bath electrons screening the impurity~\cite{Hewson97},
involves the entanglement between the impurity and the bath electrons.
This {\it impurity-bath entanglement} provides a quantum information perspective on quantum impurity systems~\cite{Bayat10,Bayat12,Bayat14,Lee15,Alkurtass16,Bayat17,Yoo17}.

For quantum impurity systems, entanglement at finite temperature can provide new information in comparison with zero-temperature entanglement of ground states.
For example, the impurity-bath entanglement exhibits power-law scaling in the Kondo regime,
and its power exponent differs between the Fermi liquid in the single-channel Kondo model and the non-Fermi liquid in the two-channel Kondo model~\cite{Lee15}.

Despite the importance,
the impurity-bath entanglement has not been computed exactly at finite temperature~\cite{Lee15} due to the following difficulty.
While pure quantum states (e.g., ground states) contain no classical correlation,
mixed states such as thermal states generally have both quantum entanglement and classical correlation~\cite{Plenio07,Guhne09,Horodecki09}.
These two different types of correlations are not easily distinguishable;
the entanglement quantification for mixed states is NP hard~\cite{Gurvits03,Gharibian10}.
For example, computation of the entanglement of formation (EoF)~\cite{Bennett96}, a mixed-state generalization of the entanglement entropy, generally requires heavy optimization. 

Therefore a practical choice of  an entanglement measure for thermal states is the entanglement negativity~\cite{Lee00,Vidal02,Plenio05}, as the negativity can be computed exactly (although it cannot detect the bound entanglement~\cite{Horodecki98}).
The negativity $\mc{N}$ between a subsystem $A$ and its complementary $B$ is
\begin{align}
\mc{N}(\rho) = \tr \, |\rho^{T_A}| - \tr \, \rho,
\label{eq:Nrho}
\end{align}
where $\rho$ is the density matrix of a target system, 
$\rho^{T_A}$ is the partial transpose of  $\rho$ with respect to the subsystem $A$,
$\tr \, |\rho^{T_A}|$ is the sum of the singular values of $\rho^{T_A}$,
and $\tr \, \rho$ is the trace of $\rho$.
To quantify the impurity-bath entanglement, one assigns $A$ the impurity and $B$ the bath.
$\mc{N}(\rho)$ is computable as long as $\tr \, |\rho^{T_A}|$ is.
Due to this computational advantage,
the negativity has been widely used to study entanglement in many-body systems at finite temperature~\cite{Audenaert02,Anders08,Eisler14,Eisler15,Eisert16,Calabrese15,Sherman16,Park17,Hart17}.

The numerical computation of the negativity $N(\rho)$, however, becomes difficult, as the size of $\rho$ becomes larger.
The difficulty appears for quantum impurity systems at finite temperature because of the following reasons.
First, the Kondo cloud~\cite{Affleck10,Park13} is a macroscopic object whose size exponentially increases with decreasing Kondo coupling strength.
Second, quantum impurity systems are generally gapless,
so their thermal density matrix involves many eigenstates and has high rank.

In this paper, we develop a numerical renormalization group (NRG)~\cite{Wilson75,Bulla08} method to compute the entanglement negativity between the impurity and the bath of quantum impurity models at finite temperature.
We construct the thermal density matrix in the complete basis set of the energy eigenstates, and then evaluate the negativity, by applying the NRG approximation, which has been originally introduced to obtain impurity correlation fuctions~\cite{Peters06,Weichselbaum07,Weichselbaum12}.

Employing the method, we compute the temperature dependence of the negativity in
the single-impurity Kondo model (SIKM) and the single-impurity Anderson model (SIAM),
the simplest models exhibiting the Kondo effect.
In the SIKM, the negativity exhibits a universal quadratic temperature dependence in the Kondo regime at low temperature, the Kondo crossover at intermediate temperature, and a sudden death~\cite{Yu09} at high temperature.
In the SIAM, both the spin and charge degrees of freedom at the impurity affect the negativity.
The impurity spin behaves in the same way as in the SIKM, while the charge fluctuation remain even at zero temperature as long as the on-site Coulomb repulsion at the impurity is finite.
To show this, we compute the negativity between the total degrees of freedom of the impurity and the bath,
and the negativity between the spin degree of freedom of the impurity and the bath.
The former depends on the Coulomb repulsion strength, and the latter shows the same quadratic scaling as in the SIKM.
Finally, we demonstrate that our method is sufficiently accurate by computing and analyzing its errors for the example of the SIKM.

This paper is organized as follows.
In \Sec{200},
we explain how to construct a thermal density matrix of an impurity problem by the NRG,
and the NRG approximation.
We apply the NRG approximation to the impurity-bath negativity and propose how to compute the negativity in Sec.~\ref{300}. 
We compute the negativity for the SIKM in Sec.~\ref{400}, and the SIAM in Sec.~\ref{500}.
We estimate and analyze the errors in our method in Sec.~\ref{600}.
Conclusion is given in Sec.~\ref{700}.

\section{Numerical Renormalization Group}\label{200}

The NRG is a powerful non-perturbative method to solve quantum impurity systems.
It provides an efficient way to construct a thermal density matrix by using a complete basis of many-body energy eigenstates~\cite{Anders05,Weichselbaum07},
over a wide range of temperature, in the thermodyamic limit.
In this section, we provide model Hamiltonians, notations, and brief introduction to the NRG including the NRG approximation.

\subsection{Model Hamiltonian}\label{210}

In this work, we apply the NRG to two paradigmatic impurity models, the SIKM and the SIAM.
The SIKM describes a spin-$1/2$ impurity interacting with the bath of conduction electrons,
\begin{equation}
H^\mr{SIKM} = J\vec{S}_\mr{d} \cdot\vec{s}_0 + \sum_{\mu} \int \mr{d}\epsilon \, \epsilon \, c^{\dagger}_{\epsilon\mu}c_{\epsilon\mu}.
\label{eq:H_Kondo}
\end{equation}
Here $J > 0$ is the coupling strength, $\vec{S}_\mr{d}$ the impurity spin,
$c_{\epsilon\mu}$ the operator annihilating a bath electron of spin $\mu = \ua, \da$ and energy $\epsilon$,
$\vec{s}_0 = \int \mr{d}\epsilon \int \mr{d}\epsilon' \sum_{\mu \mu'} c_{\epsilon\mu}^\dagger [\vec{\sigma}]_{\mu\mu'} c_{\epsilon'\mu'} / 2$
the spin of the bath electron at
the impurity site,
and $\vec{\sigma}$ the vector of the Pauli matrices.
We consider the bath of constant density of states within $[-D,D]$.
We set the half-bandwidth $D \equiv 1$ as the energy unit,
and set $\hbar = k_\mr{B} = 1$ henceforth.

On the other hand,
the SIAM contains a fermionic site with local repulsive Coulomb interaction at the impurity,
\begin{equation}
\begin{aligned}
H^\mr{SIAM} &=
\sum_{\mu} \epsilon_\mr{d} n_{\mr{d}\mu} + Un_{\mr{d}\uparrow}n_{\mr{d}\downarrow} +
\sum_{\mu} \int \mr{d}\epsilon \, \epsilon \, c^{\dagger}_{\epsilon\mu}c_{\epsilon\mu} \\
& \quad + \sum_\mu \int \mr{d} \epsilon \, \sqrt{\frac{\Gamma (\epsilon)}{\pi}} (d_{\mu}^\dagger c_{\epsilon\mu} + c_{\epsilon\mu}^\dagger d_{\mu}).
\end{aligned}
\label{eq:H_SIAM}
\end{equation}
Here $d_\mu$ annihilates a spin-$\mu$ particle at the impurity, $n_{d\mu} \equiv d_{\mu}^\dagger d_{\mu}$ is the number operator,
$\epsilon_\mr{d}$ the on-site energy at the impurity, $U$ the Coulomb interaction strength,
and $\Gamma (\epsilon)$ the hybridization function.
Throughout this work, we consider $\epsilon_\mr{d} = -U/2$ to make the impurity half-filled $\la n_{d\mu} \ra = 1/2$,
and the constant hybridization function $\Gamma (\epsilon) = \Gamma \Theta ( D- |\epsilon|)$ which relates to the constant density of states within $[-D,D]$.

Despite different type of impurities, both the SIKM and the SIAM can exhibit the Kondo effect.
It is natural since the SIKM can be derived from the SIAM as the low-energy effective Hamiltonian, via the Schrieffer-Wolff transformation~\cite{Hewson97}.

\subsection{Thermal density matrix}\label{220}

The NRG starts with the logarithmic discretization of the bath.
The bath of energy interval $[-1,1]$ is discretized by a logarithmic energy grid $\pm \Lambda^{-k+z}$ for $k = 1, 2, \cdots$,
where $\Lambda > 1$ is a discretization parameter and $z = 0, \tfrac{1}{n_z} \cdots, 1 - \tfrac{1}{n_z}$ is the discretization shift~\cite{Oliverira94,Campo05}.
Then the impurity model is mapped onto the so-called Wilson chain
where the bath degrees of freedom lie along a tight-binding chain and the impurity couples to one end of the chain.
The models in Eqs.~\eqref{eq:H_Kondo} and \eqref{eq:H_SIAM} are mapped onto
the chain Hamiltonians,
\begin{align}
H^\mr{SIKM}_N &= J \vec{S}_\mr{d} \cdot \vec{s}_0 + H_N^\mr{bath}, \label{eq:HN_SIKM}\\
H^\mr{SIAM}_N &= \sum_{\mu} \epsilon_\mr{d} n_{\mr{d}\mu} + Un_{\mr{d}\uparrow}n_{\mr{d}\downarrow} + H_N^\mr{bath} \nonumber\\
&\quad + \sqrt{\frac{2\Gamma}{\pi}} \sum_{\mu} (d_{\mu}^\dagger f_{0\mu} + f_{0\mu}^\dagger d_{\mu}), \label{eq:HN_SIAM}
\end{align}
where
$H^\mr{bath}_N = \sum_\mu \sum_{n = 1}^N t_{n} f_{n-1,\mu}^\dagger f_{n\mu} + \text{H.c.}$ is the bath Hamiltonian of the chain length $N+1$,
$f_{n\mu}$ annihilates a spin-$\mu$ particle at site $n \in [0,N]$,
and $\vec{s}_0$ is the spin operator at site $0$ next to the impurity.
Due to the logarithmic discretization, the hopping amplitudes decay exponentially as $t_n \sim \Lambda^{-n/2}$.
In practice, we consider the chain of a finite $N$ such that its lowest energy scale $\sim \Lambda^{-N/2}$ is smaller than any other physical energy scales such as the system temperature $T$.

The Fock space of the chain is spanned by the basis
$\{ \ket{s_\mr{d}} \otimes \ket{s_0} \otimes \cdots \otimes \ket{s_N} \}$,
where $\ket{s_\mr{d}}$ is the impurity state and $\ket{s_n}$ is the state of a bath site $n$.
Since the Fock space dimension of the chain scales as $O (d^N)$ 
(here $d = 4$ is the dimension of each bath site for the single-channel problems considered in this work),
it is hard to exactly diagonalize the chain with large $N$.

By taking advantage of the exponential decay of the hopping amplitudes,
one can construct the complete basis of the energy eigenstates by using the iterative diagonalization~\cite{Anders05,Peters06}.
In the $n$th iterative diagonalization step, one obtains a set of energy eigenstates in an energy window $[E_{n 1}^K, E_{n i_\textrm{max}}^D]$ for a short chain composed of sites from the impurity to site $n$, where $E_{n 1}^K$ and $E_{n i_\textrm{max}}^D$ are the lowest and highest energies of the set.
The energy level spacing between these eigenstates is of the order of $t_n \sim \Lambda^{-n/2}$. 
Then, one separates the set into two subsets, the ``discarded'' energy eigenstates $\{ \ket{E_{n i}^D} \}$ and the ``kept'' eigenstates $\{ \ket{E_{n i}^K} \}$, by energy.
Here these eigenstates are indexed by a common index $i$ such that their corresponding energy eigenvalues are in increasing order; 
the kept states are within energy window $[E_{n 1}^K, E_{n N_{\mr{tr}}}^K]$, while the discarded states are in $[E_{n, N_\mr{tr} + 1}^D, E_{n,i_{\max}}^D ]$,
where $N_\mr{tr}$ is the number of the kept states
and $i_{\max}$ is the number of total states at a given iteration $n$.
One typically takes $E_\mr{tr} \equiv (E_{n, N_\mr{tr}+1}^D - E_{n, 1}^K)/\Lambda^{-n/2} \gtrsim 7$~\cite{Weichselbaum12}.
In the $(n+1)$th diagonalization step, one constructs 
the Hilbert space $\{ \ket{E_{n i}^K} \otimes \ket{s_{n+1}} \}$ and diagonalize the Hamiltonian for a longer chain composed of the short chain and the next site $n+1$.
One iterates these processes until one reaches the last site $N$.
At the last iteration, all the eigenstates are discarded.

The discarded states $\{ \ket{E_{ni}^D} \}$ decouple from the states of the sites $n' > n$, $\{ \ket{s_{n+1}} \otimes \cdots \otimes \ket{s_{N}} \}$, which we call the environment states of $\{ \ket{E_{ni}^D} \}$. 
The whole Fock space can be constructed by the complete basis states of
\begin{equation}
\Big\{ \ket{E_{ni\vec{s}}^D} \equiv \ket{E_{ni}^D} \otimes \ket{s_{n+1}} \otimes \cdots \otimes \ket{s_N} \Big| n=n_0, n_0 + 1, \cdots N \Big\},
\label{eq:CompBasis}
\end{equation}
where $n_0$ is the earliest iteration at which the Hilbert space truncation happens.
These basis states can be used as the approximate eigenstates of the full Hamiltonian (the whole Wilson chain),
and $E_{ni}^D$ provides an approximate eigenenergy.
Based on energy scale separation,
the approximation error $\delta E_{ni}^D$ for each energy $E_{ni}^D$,
which originates from neglecting its coupling to the environment states, is estimated by $\delta E_{ni}^D / E_{ni}^D \sim t_{n+1} / E_{n,N_\mr{tr}+1}^D \sim 1 / E_\mr{tr} \sqrt{\Lambda} \ll 1$.
Therefore, for large enough $\Lambda$ and $E_\mr{tr}$,
the basis states in Eq.~\eqref{eq:CompBasis} are efficient
description of energy eigenstates,
since the total number $O (N_\mr{tr} N)$ of $\{ |E_{ni}^D \rangle \}$ is much smaller than $O (d^N)$. 

Using the complete basis states in \Eq{eq:CompBasis},
one writes the thermal density matrix $\rho_T$ at temperature $T$ as
\begin{gather}
\begin{aligned}
\rho_T &= \sum_{n=n_0}^N\sum_{i\vec{s}} \frac{e^{-E_{ni}^D / T}}{Z} \prj{E_{ni\vec{s}}^D} = \sum_{n=n_0}^N \mc{R}_n,
\end{aligned} \label{eq:rho_T} \\
\mc{R}_n \equiv \rho_n^D \otimes I_{n+1} \otimes \cdots \otimes I_{N}, \label{eq:R_n} \\
\rho_n^D = \sum_i \frac{d^{N-n} e^{-E_{ni}^D/T}}{Z} \prj{E_{ni}^D},  \label{eq:rho_n^D}
\end{gather}
where $I_n = \sum_{s_n} \prj{s_n} / d$ is the identity
with normalization $\tr \, I_n = 1$,
and $Z$ is the partition function.

\subsection{NRG approximation of correlation functions}\label{230}

The complete basis $\{ \ket{E_{ni\vec{s}}^D} \}$ provides the systematic way of computing various physical properties.
One needs to use the NRG approximation~\cite{Weichselbaum07,Weichselbaum12}, to reduce the cost of computing
matrix elements $\matel{E_{ni\vec{s}}^D}{\mc{O}}{E_{n'i'\vec{s}'}^D}$ of an operator $\mc{O}$.
Since we will apply the NRG approximation to compute negativity in \Sec{300},
we here briefly explain the NRG approximation for computing the impurity correlation function.

By using the complete basis, the impurity correlation function can be expressed in the Lehmann representation
\begin{equation}
\begin{aligned}
\mc{A} (\omega) &\equiv \frac{1}{\pi} \mr{Im} \int_{-\infty}^{\infty} \mr{d} t \, e^{i\omega t} i \Theta (t) \tr \left(\rho_T [ \mc{O} (t), \mc{O}^\dagger ]_\pm \right) \\
&= \sum_{n n' i i' \vec{s} \vec{s}'} A_{(n i \vec{s}),(n' i' \vec{s}')} \, \delta (\omega - \omega_{(ni),(n'i')}),
\label{eq:A_full}
\end{aligned}
\end{equation}
\begin{equation}
\begin{aligned}
A_{(n i \vec{s}),(n' i' \vec{s}')} &= | \matel{E_{ni\vec{s}}^D}{\mc{O}}{E_{n'i'\vec{s}'}^D} |^2 ( \rho_{n i \vec{s}} \pm \rho_{n'i'\vec{s}'}), \\
\rho_{n i \vec{s}} &= \epv{E_{ni\vec{s}}^D}{\rho_T} =  e^{- E_{ni}^D / T} / Z , \\
\omega_{(ni),(n'i')} &= E_{n'i'}^D - E_{ni}^D,
\end{aligned}
\label{eq:A_full2}
\end{equation}
where $\mc{O}$ is the local operator acting on the impurity and $+$ ($-$) in $\pm$ is for a fermionic (bosonic) operator $\mc{O}$.

Direct calculation of Eq.~\eqref{eq:A_full} is impractical, since the number of matrix elements $A_{(n i \vec{s}),(n' i' \vec{s}')}$ is $O(N_\textrm{tr}^2 d^{2N})$. 
To make the calculation feasible, one applies the NRG appriximation, with which the number is significantly reduced to  $O(N_\textrm{tr}^2 N)$. 
The approximation is accurate within the intrisic error of the NRG that the inaccuracy of the energies  $E_{ni\vec{s}}^D$ is estimated as $\delta E^D_n \sim \Lambda^{-(n+1)/2}$. 

We now explain the NRG approximation.
In the calculation of Eq.~\eqref{eq:A_full}, one applies the identity of $\sum_{n' >n; i \vec{s}} | E^D_{n' i \vec{s}} \rangle \langle E^D_{n' i \vec{s}} |= \sum_{i' \vec{s}'} |E^K_{n i' \vec{s}'} \rangle \langle E^K_{n i' \vec{s}'}|$ and approximately treats
$|E^K_{n i' \vec{s}'} \rangle$ as an eigenstate of the full Hamiltonian, although $|E^K_{n i' \vec{s}'} \rangle$ is an eigenstate of the NRG chain with incomplete chain length $n+1$.
As a result,
an energy differences $\omega_{(ni),(n'i')} = E_{n'i'}^D - E_{ni}^D$ is replaced by 
$E_{n i''}^K - E_{ni}^D$ if $n' > n$ or by $E_{n'i'}^D - E_{n'i''}^K$ if $n' < n$.
The error in $\omega_{(ni),(n'i')}$, i.e., $\delta \omega_{(ni),(n'i')} \sim \Lambda^{-(n+1)/2}$
due to this replacement, is comparable with the error of the Hilbert space truncation $\sim \delta E_{ni}^D \sim \Lambda^{-(n+1)/2}$.

The NRG approximation simplifies the summation in \Eq{eq:A_full} without inducing further numerical error:
Only the matrix elements $ \matel{E_{ni\vec{s}}^X}{\mc{O}}{E_{n i'\vec{s}'}^{X'}}$ diagonal in $n$ remain in the subsequent steps as $\matel{E_{ni}^X}{\mc{O}}{E_{n i'}^{X'}} \delta_{\vec{s} \vec{s}'}$,
which removes the sum over $\sum_{\vec{s}\vec{s}'}$ and reduces the computation cost to $O(N_\textrm{tr}^2 N)$ mentioned above.
Then $\mc{A} (\omega)$ becomes
\begin{equation}
\mc{A} (\omega) \approx \sum_{n X X' i i'}^{(X, X') \neq (K, K)} \tilde{A}_{n,(Xi) ,(X'i')} \, \delta (\omega - \tilde{\omega}_{n,(Xi),(X'i')}),
\label{eq:A_app1}
\end{equation}
\begin{equation}
\begin{aligned}
\tilde{A}_{n,(Xi) ,(X'i')} &= | \matel{E_{ni}^X}{\mc{O}}{E_{ni'}^{X'}} |^2 ( \rho_{n i}^X \pm \rho_{n i'}^{X'}), \\
\rho_{n i}^X &= \epv{E_{ni}^X}{\rho_n^X} , \\
\tilde{\omega}_{n,(Xi),(X'i')} &= E_{ni'}^{X'} - E_{ni}^X ,
\end{aligned}
\label{eq:A_app2}
\end{equation}
where $(X,X') = (D,D)$, $(D,K)$, or $(K,D)$;
the case $(X,X') = (K,K)$ is excluded to avoid double counting.
The density matrix $\rho_n^K$ 
\begin{align}
\rho_n^K &= \tr_{n+1, \cdots, N}  \Bigg[ \sum_{n' > n}^N \mathcal{R}_{n'} \Bigg],
\end{align}
is introduced in the calculation;
$\mathcal{R}_{n'}$ is defined in \Eq{eq:R_n} and
$\tr_{n+1, \cdots, N} (\cdot) \equiv 
\sum_{s_{n+1}, \cdots, s_{N}} \bra{s_{N}} \otimes \cdots \otimes \bra{s_{n+1}} (\cdot) \ket{s_{n+1}} \otimes \cdots \otimes \ket{s_{N}}$.

Summarizing, consider a contribution to the spectral function, which involves the eigenstates $\ket{E_{ni\vec{s}}^D}$ and $\ket{E_{n'i'\vec{s}'}^D}$ from different iterations $n$ and $n' (> n)$.
The NRG approximation neglects the detailed information of the later sites $n' > n$ by tracing them out.
Then the contribution is simplified to the approximated one involving the discarded and kept states at the {\it same} iteration,
say $\ket{E_{ni}^D}$ and $\ket{E_{ni'}^K}$.
As long as the energy scale separation $1/\sqrt{\Lambda}E_\mr{tr} \ll 1$ holds by appropriately choosing parameters ($\Lambda$, $N_\mr{tr}$, and/or $E_\mr{tr}$),
the result obtained after the NRG approximation is accurate;
for example, the impurity spectral function at $\omega = 0$ and $T = 0$ satisfies the Friedel sum rule within sub-1\% error~\cite{Weichselbaum07}.

The NRG approximation is equivalent to 
the replacements of $\mathcal{R}_n$ by $\rho_n^D$ and $\sum_{n' > n}^N \mathcal{R}_{n'}$ by $\rho_n^K$ in the calculation,
\begin{equation}
\begin{gathered}
\mathcal{R}_n = \rho_n^D \otimes I_{n+1} \otimes \cdots \otimes I_{N} \to \rho_n^D,
\\
\sum_{n' > n}^N \mathcal{R}_{n'} = \sum_{n' > n}^N \rho_{n'}^D \otimes I_{n'+1} \otimes \cdots \otimes I_{N} \to \rho_n^K.
\label{NRGAPP} 
\end{gathered}
\end{equation}
Here, the information of sites $n' > n$ is traced out.
This is in parallel to that 
$|E^D_{n i \vec{s}} \rangle$ and $|E^K_{n i \vec{s}} \rangle$ are approximately treated as an eigenstate of the full Hamiltonian.
We apply these replacements for computing $\mc{N}$ below.

\section{NRG method for negativity}\label{300}

We propose how to compute the negativity $\mc{N}$ in \Eq{eq:Nrho} that quantifies the impurity-bath entanglement of the thermal density matrix $\rho_T$ in \Eq{eq:rho_T}.
$\mc{N} (\rho_T)$ is computed in two steps,
taking partial transpose on $\rho_T$ to get $\rho^{T_A}_T$ and then diagonalizing $\rho^{T_A}_T$.
However, one cannot compute $\mc{N}$ directly applying these two steps, since
the environment states $I_{n+1} \otimes \cdots \otimes I_{N}$ in \Eq{eq:rho_T} make the dimension of $\rho_T$ exponentially large $\sim O(d^N)$.
We overcome this difficulty by utilizing the NRG approximation.

\subsection{NRG approximation of negativity}\label{310}

To start with, we decompose the expression of $\mc{N} (\rho_T)$.
\begin{gather}
\mc{N}(\rho_T) = \mc{N} \bigg( \sum_{n=n_0}^N  \mathcal{R}_n \bigg) = \sum_{n=n_0}^N \mc{N} (\mathcal{R}_n) - \sum_{n=n_0}^N \delta_n,
\label{eq:Nexact} \\
\delta_n \equiv \mc{N} (\mathcal{R}_n) + \mc{N} \bigg( \sum_{n' > n}^N \mathcal{R}_{n'} \bigg) - \mc{N} \bigg( \mathcal{R}_n + \sum_{n' > n}^N \mathcal{R}_{n'} \bigg).
\label{eq:dexact}
\end{gather}
In \Eq{eq:Nexact}, $\mc{N}(\rho_T)$ has two parts, $\sum_n \mc{N}(\mathcal{R}_n)$ and $\sum_n \delta_n$.
The first part $\sum_n \mc{N}(\mathcal{R}_n)$ is the sum of the entanglement in each density matrix $\mathcal{R}_n$, and the second $\sum_n \delta_n$ counts contribution from mixtures of different $\mathcal{R}_n$'s.
Due to the convexity of the negativity~\cite{Vidal02, Plenio05}, $\delta_n \geq 0$ is guaranteed.
Equations~\eqref{eq:Nexact} and \eqref{eq:dexact} are exact,
given construction of density matrix $\rho_T$.

One can derive the expression in \Eq{eq:Nexact}, applying the definition of $\delta_n$ in \Eq{eq:dexact} 
recursively:
(i) Start from the iteration step  $n_0$ at which the first Hilbert space truncation happens during the iterative diagonalization.
Using Eq.~\eqref{eq:rho_T} and the definition of $\delta_{n = n_0}$, one decomposes the negativity $\mc{N}(\rho_T)$ as
\begin{equation}
\mc{N}(\rho_T) = \mc{N} \left( \mc{R}_{n_0} \right) - \delta_{n_0} + \mc{N} \left( \sum_{n' > n_0}^{N} \mc{R}_{n'} \right) .
\label{eq:Ninit}
\end{equation}
(ii) Next, we use an inductive argument. Suppose that one can decompose the negativity $\mc{N}(\rho_T)$ as
\begin{equation}
\mc{N}(\rho_T) = \sum_{n' = n_0}^n \mc{N} (\mathcal{R}_{n'}) - \sum_{n'=n_0}^n \delta_{n'}  + \mc{N} \bigg(\sum_{n' > n}^N \mathcal{R}_{n'} \bigg) .
\label{eq:Nrecur}
\end{equation}
Then, one decomposes \Eq{eq:Nrecur} by rewriting the last term in its right hand side using $\delta_{n+1}$ (cf.~\Eq{eq:dexact}).
\begin{equation}
\mc{N}(\rho_T) = \sum_{n' = n_0}^{n+1} \mc{N} (\mathcal{R}_{n'}) - \sum_{n'=n_0}^{n+1} \delta_{n'}  + \mc{N} \bigg(\sum_{n' > n+1}^N \mathcal{R}_{n'} \bigg) .
\label{eq:Nrecur2}
\end{equation}
Notice that
\Eq{eq:Nrecur2} remains in the same form as the index $n$ increases to $n+1$. 
By induction, one obtains \Eq{eq:Nexact}.

Now we apply the NRG approximation to compute $\delta_n$.
The second and third terms on the right hand side of \Eq{eq:dexact} involve the density matrices $\mc{R}_{n'}$ from different iterations $n' (> n)$.
As done in the correlation functions (see \Sec{230} or  \Eq{NRGAPP}),
we trace out the later sites $n' > n$ for the arguments $\sum_{n'>n}^N \mc{R}_{n'}$ and $\mc{R}_n + \sum_{n'>n}^N \mc{R}_{n'}$.
Accordingly we have
\begin{align}
\mathcal{N}(\mathcal{R}_n) &\to \mathcal{N}(\rho_n^D),\\
\delta_n &\to \mc{N}( \rho_n^D ) + \mc{N} (\rho_n^K) - \mc{N}( \rho_n^D + \rho_n^K) \equiv \delta_n^{[0]}.
\label{eq:deltB00}
\end{align}
The superscript $[0]$ indicates that the NRG approximation is applied to $\delta_n$.
Then, the negativity $\mc{N}(\rho_T)$ is computed using $\mc{N}(\rho_n^D)$, $\mc{N}(\rho_n^K)$, and $\mc{N}(\rho_n^D + \rho_n^K)$.

The dimension of the matrices $\rho_n^D$, $\rho_n^K$, $\rho_n^D + \rho_n^K$ is independent of $N$
and less than or equal to $O(d N_\textrm{tr})$,
which is exponentially smaller than the dimension $\sim O(d^N)$ of $\rho_T$.
This reduction of the matrix size makes  computation of $\mathcal{N}$ feasible.
As we will discuss in \Sec{600}, the error generated by the NRG approximation in \Eq{eq:deltB00}
is smaller than or comparable to the intrinsic error of the NRG in computing $\mathcal{N}$.

\subsection{Constructing impurity-bath bipartite basis}\label{320}

To compute $\mc{N}(\rho_n^D)$, $\mc{N}(\rho_n^K)$, and $\mc{N}(\rho_n^D + \rho_n^K)$,
one needs to represent the eigenstates $\{ \ket{E_{ni}^X} \}$ $(X=D,K)$ 
in the bipartite basis of the impurity and the bath as
\begin{align}
\ket{E_{ni}^X} \equiv \sum_{j, s_\mr{d}} [T_n^{X}]_{s_\mr{d},j,i} \ket{s_\mr{d}} \otimes \ket{\phi_{nj}}.
\label{EQ18}
\end{align}
Here $\ket{s_\mr{d}}$ is the impurity state, 
$\ket{\phi_{nj}}$ is the bath state satisfying $\ket{\phi_{nj}} \in \mr{span}\{ \ket{s_0} \otimes \cdots \otimes \ket{s_n} \}$, 
$\ovl{\phi_{nj}}{\phi_{nj'}} = \delta_{jj'}$,
and $T_n^X$ is the ``coefficient'' tensor whose element is
\begin{equation}
[T_n^{X}]_{s_\mr{d},j,i} = \left( \bra{s_\mr{d}} \otimes \bra{\phi_{nj}} \right) \ket{E_{ni}^X}.
\end{equation}
Given coefficient tensor $T_n^X$,
we express the states $\rho_n^X$ in the basis of $\{ \ket{s_\mr{d}} \otimes \ket{\phi_{nj}} \}$,
to take the partial transpose $(\rho_n^X)^{T_A}$ with respect to $\{ \ket{s_\mr{d}} \}$.
Then we evaluate $\tr \, | (\rho_n^X)^{T_A} |$ by obtaining the singular value decomposition 
(or equivalently, eigendecomposition)
of $(\rho_n^X)^{T_A}$.
It is the same for the sum $\rho_n^D + \rho_n^K$.

We iteratively construct $T_n^X$ from $T_{n-1}^{K}$ and $U_n^X$,
where $U_n^X$ is a left-unitary matrix which relates the eigenstates at iterations $n-1$ and $n$,
\begin{equation}
\begin{gathered}
\phantom{.} [U_n^X]_{s_n,k,i} \equiv \left( \bra{s_n} \otimes \bra{E_{n-1,k}^K} \right) \ket{E_{ni}^X},
\\
\sum_{s_n,k} [U_n^{X}]^*_{s_n,k,i} [U_n^{X'}]_{s_n,k,i'} = \delta_{XX'} \delta_{ii'} ,
\end{gathered}
\end{equation}
where $X, X' = D, K$.
We construct these matrices $T_n^X$ and $U_n^X$ during the standard NRG iterative diagonalization. 

We start the iterative construction from $T_0^X$ with the bath state $\ket{\phi_{0,j=s_0}} \equiv \ket{s_{0}}$,
\begin{equation}
[T_0^{X}]_{s_\mr{d},s_0,i} \equiv \left( \bra{s_\mr{d}} \otimes \bra{s_0} \right) \ket{E_{0i}^X}.
\end{equation}
Then consider an iteration $n$, and suppose we know $T_{n-1}^K$ at the earlier iteration $n-1$.
We first obtain $U_n^X$ which diagonalizes the Hamiltonian at the current iteration $n$.
Then we construct the matrix $Q_n^X$ in terms of $T_{n-1}^{K}$ and $U_n^X$ as
\begin{equation}
\begin{aligned}
\phantom{.}[Q_n^X]_{(j',s_n),(s_\mr{d},i)} &\equiv 
\left( \bra{s_\mr{d}} \otimes \bra{\phi_{n-1,j'}} \otimes \bra{s_n} \right) \ket{E_{ni}^X} \\
&= \sum_{k} [T_{n-1}^{K}]_{s_\mr{d},j',k} [U_n^X]_{s_n, k, i}.
\end{aligned}
\end{equation}
To ensure the orthonormality of $\{ \ket{\phi_{nj}} \}$, 
we perform the singular value decomposition as
\begin{equation}
[Q_n^K + Q_n^D]_{(j', s_n), (s_\mr{d}, i)} = \sum_{j} [V_L]_{(j', s_n), j} [\Sigma V_R^\dagger]_{j, (s_\mr{d}, i)},
\label{eq:Q_SVD}
\end{equation}
where $V_L$ and $V_R$ are unitary matrices, $\Sigma$ is the diagonal matrix of non-zero singular values, and $Q_n^K$ and $Q_n^D$ act on disjoint set of column indices $(s_\mr{d}, i)$.
Based on its unitarity, we assign $V_L$ as the matrix which defines the mapping from $\{ |\phi_{nj} \rangle \}$ to $\{| \phi_{n-1,j'} \rangle \otimes | s_n \rangle \}$ such that $[V_L]_{(j', s_n), j} = ( \bra{\phi_{n-1,j'}} \otimes \bra{s_n} ) |\phi_{nj} \rangle$.
Hence we construct the desired tensor $T_n^X$,
\begin{align}
[T_n^{X}]_{s_\mr{d},j,i} &= \sum_{j',s_n} [V_L]^*_{j,(j',s_n)} [Q_n^X]_{(j',s_n),(s_\mr{d},i)} .
\label{EQ23}
\end{align}
Note that $V_L$ is left-unitary; the multiplication of non-square $V_L^\dagger$ in \Eq{EQ23} indicates the truncation of the bath Hilbert space.

After this iterative construction,
the dimension of the bath space spanned by $\{ \ket{\phi_{nj}} \}$ for a single $n$ scales as $O (d_\mr{imp} N_\mr{tr})$;
the maximum number of non-zero singular values in the decomposition of \Eq{eq:Q_SVD} is $O(d_\mr{imp} N_\mr{tr})$.
Thus the matrix form of $\rho_n^D + \rho_n^K$ in the basis of $\{ \ket{s_\mr{d}} \otimes \ket{\phi_{nj}} \}$
has dimension $O (d_\mr{imp}^2 N_\mr{tr})$.
The computational cost of evaluating the singular value decomposition of $(\rho_n^D + \rho_n^K)^{T_A}$,
which is the most computationally demanding part in computing the negativity,
is the cube of the matrix dimension, i.e., $O (d_\mr{imp}^6 N_\mr{tr}^3)$.

This estimation indicates that the cost of computing the negativity for the SIAM ($d_\mr{imp} = 4$) will be $64$ times larger than that for the SIKM ($d_\mr{imp} = 2$)
if the other numerical parameters are the same.

\subsection{Symmetry}\label{330}

Quantum impurity systems possess various symmetries such as $\mr{U}(1)$ charge symmetry and $\mr{SU}(2)$ spin symmetry.
The NRG exploits these symmetries to reduce the computational cost and to increase the numerical accuracy~\cite{Toth08,Weichselbaum12,Weichselbaum2012:sym}.
For example, a thermal density matrix $\rho_T$ possesses the symmetries of its Hamiltonian, hence, it can be computed and represented efficiently in a block diagonal form whose blocks are labelled by the eigenvalues of the operators corresponding to the symmetries.
 
Unfortunately however, 
the symmetries cannot be fully exploited in computing the negativity.
Partial transpose can destroy the block diagonal form of the thermal density matrix $\rho_T$;
that is, a symmetry operator $Q$ satisfying $[Q, H] = 0$ commutes with $\rho_T$, but not necessarily with $\rho_T^{T_A}$.
For example,
the SIKM has $\mr{U}(1) \times \mr{U}(1)$ symmetry conserving spin-up charge (the corresponding symmetry operator is the spin-up particle number operator $Q_\uparrow$) and spin-down charge ($Q_\downarrow$).
Consider a nonzero matrix element $\rho_{(\Uparrow\phi), (\Downarrow\phi')}$ of a density matrix $\rho$,
where $\ket{{\Uparrow}}$ and $\ket{{\Downarrow}}$ are impurity spin states.
Both $\ket{{\Uparrow}} \otimes \ket{\phi}$ and $\ket{{\Downarrow}} \otimes \ket{\phi'}$ have the same eigenvalues $(q_\uparrow, q_\downarrow)$ of $(Q_\uparrow, Q_\downarrow)$.
After partial transpose,
the matrix element
$\rho_{(\Uparrow\phi), (\Downarrow\phi')}$ 
is relocated to the position indexed by ${(\Downarrow\phi), (\Uparrow\phi')}$,
where $\ket{{\Downarrow}} \otimes \ket{\phi}$ has an eigenvalues $(q_\uparrow-1, q_\downarrow+1)$ and 
$\ket{{\Uparrow}} \otimes \ket{\phi'}$ has an eigenvalues $(q_\uparrow+1, q_\downarrow-1)$.
Therefore, to make $\rho_T^{T_A}$ block-diagonal,
one should resort to the weaker symmetry, i.e., the total charge conservation, leading to larger block size.
Even worse, for the SIAM, $\rho_T^{T_A}$ does not respect even the total charge conservation,
since the partial transpose on the impurity Hilbert space mixes up the blocks with different charges.

Since Hamiltonian symmetries may not be useful for computing $\rho_T^{T_A}$,
we choose small $N_\mr{tr} \gtrsim 100$ to treat the SIKM and the SIAM within a practical cost.
We choose large $\Lambda = 10$ to ensure energy scale separation with this small $N_\mr{tr}$.
Such large $\Lambda = 10$ can yield accurate values of static, i.e., frequency-independent quantities;
for example, impurity contributions, obtained with $\Lambda = 10$, to magnetic susceptibility or to specific heat agree with the Bethe ansatz result within a few $\%$~\cite{Merker12}.
We will show in \Sec{600} that
our result of the negativity, obtained with small $N_\mr{tr} \gtrsim 100$ and large $\Lambda = 10$, is also sufficiently accurate.

\section{Negativity in the Kondo Model} \label{400}

\begin{figure}
\centerline{\includegraphics[width=0.5\textwidth]{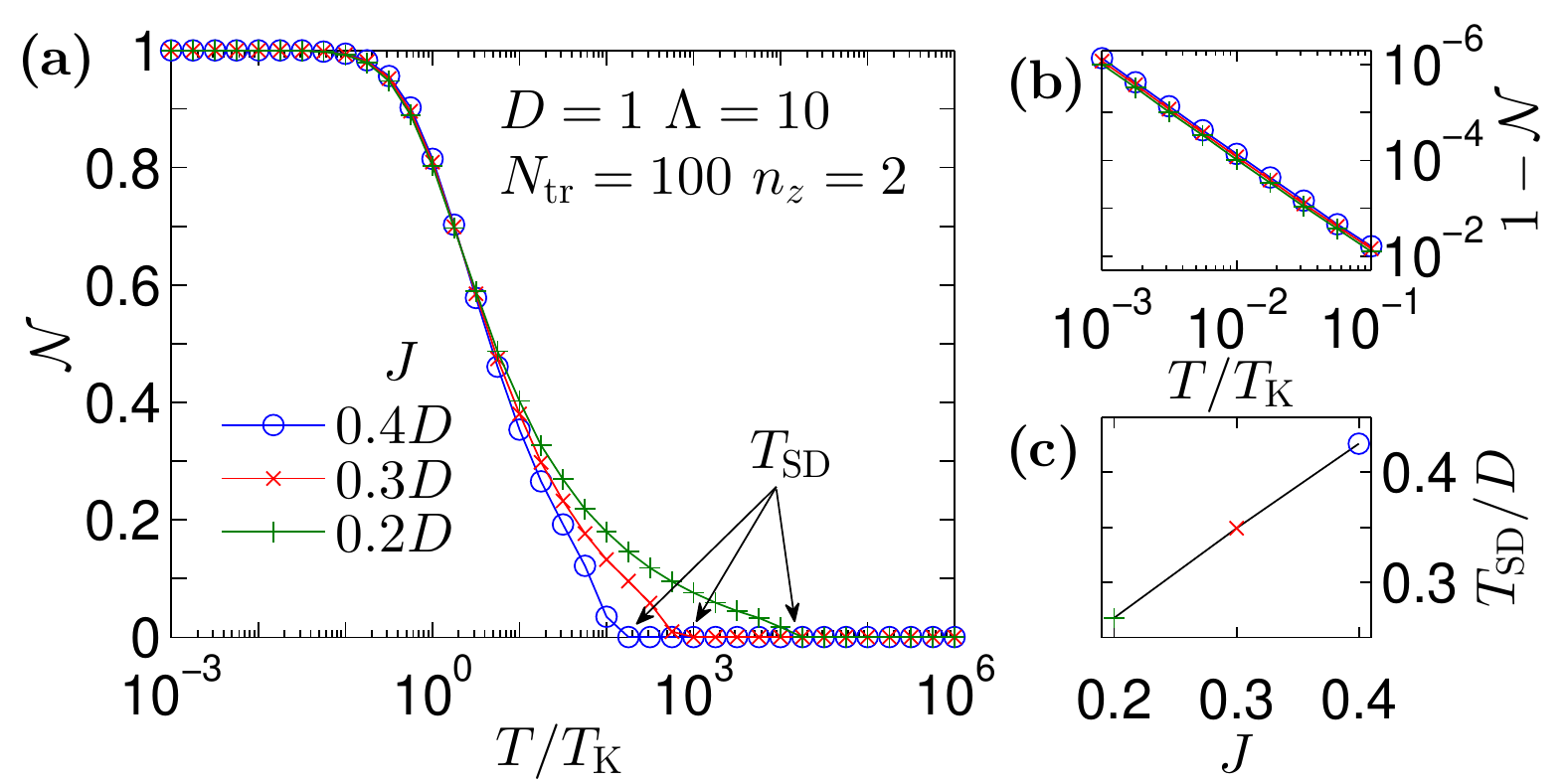}}
\caption{(Color online)
(a) Temperature ($T$) dependence of the impurity-bath negativity $\mathcal{N}$ in the SIKM
for different $J$'s.
The negativity has the maximum value $1$ at $T = 0^+$,
exhibits crossover around $T = T_\mr{K}$,
and vanishes (sudden death) at $T = T_\mr{SD} \gg T_\mr{K}$.
(b) The negativity $\mc{N} (T)$ in the Kondo regime $T \ll T_\mr{K}$.
It follows the power law of $1 - \mc{N} \sim (T / T_\mr{K})^2$ [cf.~\Eq{eq:Nscaling}].
(c) Sudden death temperature $T_\mr{SD}$ for different $J$'s.
$T_\mr{SD}$ increases linearly with increasing $J$.
}
\label{fig01}
\end{figure}

We apply the method developed in the previous section to the SIKM.
In \Fig{fig01}, we compute the temperature dependence of the negativity $\mc{N}$ that quantifies the impurity-bath entanglement in the SIKM.
The negativity $\mc{N}$ exhibits a universal Kondo behavior at low temperature $T \ll T_\mr{K}$,
shows a thermal crossover around $T = T_\mr{K}$,
and vanishes at high temperature $T \gg T_\mr{K}$.
Here the Kondo temperature is defined as
$T_\mr{K}= \sqrt{J/2D}e^{-2D/J}$.

We first explain the universal behavior of the negativity $\mc{N}$ at low temperature $T \lesssim T_\mr{K}$.
The curves $\mc{N} (T / T_\mr{K})$ of different $J$'s lie on top of each other.
At the strong-coupling fixed point of $T = 0^+$,
the impurity and the bath are entangled to form the Kondo spin singlet, as indicated by
the maximal negativity $\mc{N}=1$.
At $T \ll T_\mr{K}$, the negativity $\mc{N}$ follows the power-law scaling
\begin{equation}
\mc{N} \simeq 1 - a_{\mc{N},\mr{1CK}} (T/T_\mathrm{K})^2 ,
\label{eq:Nscaling}
\end{equation}
where a coefficient $a_{\mc{N},\mr{1CK}} > 0$ is order $O(1)$,
as shown in \Fig{fig01}(b).
This quadratic dependence originates from the low-energy excitation of the Fermi-liquid quasiparticles~\cite{Lee15},
which can be confirmed by using the bosonization. (See \App{A00} for the details.)
The behavior of the negativity $\mc{N}$ at $T \lesssim T_\mr{K}$ is consistent with that of the EoF~\cite{Lee15} quantifying the impurity-bath entanglement in the SIKM.

Next we explain the behavior of the negativity $\mc{N}$ at high temperature $T \gtrsim T_\mr{K}$.
As $T$ increases from $0^+$,
the negativity $\mc{N}$ exhibits the thermal crossover around Kondo temperature $T_\mr{K}$.
At high temperature $T \gg T_\mr{K}$, the impurity and the bath are weakly correlated, having small negativity $\mc{N} \ll 1$ at the local-moment fixed point.
The negativity $\mc{N}$ suffers sudden death~\cite{Yu09} (within numerical noise) at $T = T_\mr{SD} \sim J$ [see \Fig{fig01}(c)], that is, $\mc{N}$ is finite at $T < T_\mr{SD}$, while it vanishes at $T \geq T_\mr{SD}$.

One can understand the linear dependence of $T_\mr{SD}$ vs.~$J$ from a minimal model $H_{N=0}^\mr{SIKM}$ [see \Eq{eq:HN_SIKM}].
$H_{N=0}^\mr{SIKM}$ is composed of the impurity and only the nearest bath site,
which describes the $T \to \infty$ limit of the Wilson chain since the effective chain length scales as $\sim -2 \log_\Lambda T$~\cite{Weichselbaum07,Weichselbaum12}.
We analytically show in \App{B00} that
the minimal model $H_{N=0}^\mr{SIKM}$ exhibits the entanglement sudden death in terms of both the negativity and the EoF at $T = J / \ln 3$.
This provides the underlying mechanism of the linear dependence of $T_\mr{SD}$ vs.~$J$.
Note that the entanglement sudden death also appears in other many-body systems at finite temperature~\cite{Sherman16,Park17,Hart17}.

\section{Negativity in the Anderson Model}\label{500}

\begin{figure}
\centerline{\includegraphics[width=0.5\textwidth]{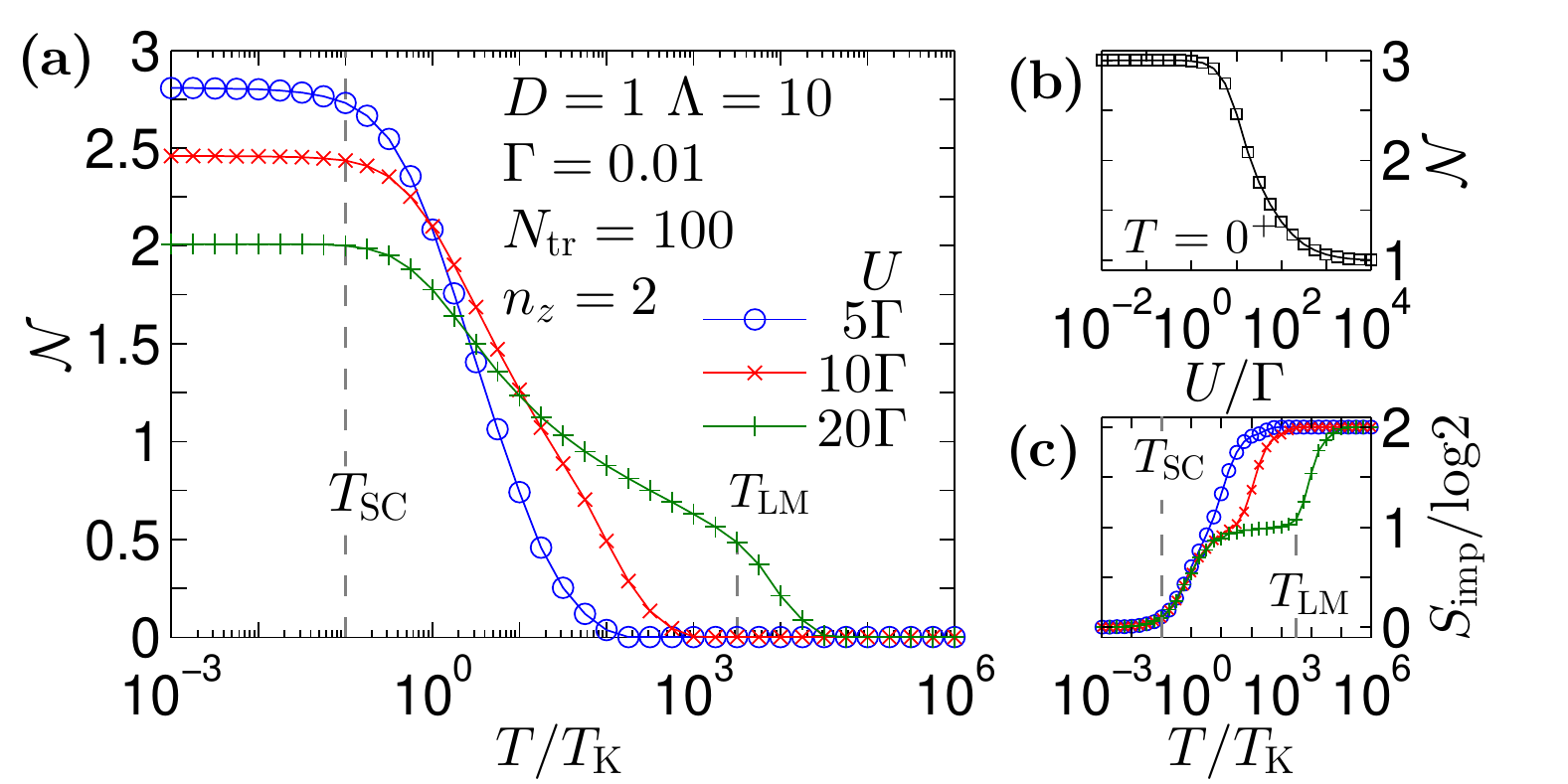}}
\caption{(Color online)
(a) Temperature ($T$) dependence of the negativity $\mc{N}$ quantifying the impurity-bath entanglement in the SIAM.
The zero-temperature values $\mc{N}(T = 0^+)$ depend on $U$.
The negativity $\mc{N}$ exhibits crossovers between different fixed points as kinks;
$\mc{N}$ shows a kink around $T = T_\mr{SC}$ for all values of $U$,
and another kink around $T = T_\mr{LM}$ for large $U = 20 \Gamma$.
(b) $\mc{N} (T = 0^+)$ decreases with increasing $U$.
(c) Impurity entropy $S_\mr{imp}$ shows the crossovers correponding to those of  $\mc{N}$.
The temperature scales $T_\mr{SC}$ and $T_\mr{LM}$ are located at the end of the plateaus in $S_\mr{imp}$,
while the plateaus indicate fixed points.
We use the Kondo temperature $T_\mr{K} = (e^{\gamma+1/4}/\pi^{3/2})\sqrt{U\Gamma/2} e^{-\pi U/8\Gamma+\pi\Gamma/2U}$ \cite{Hewson97},
where $\gamma \simeq 0.5772$ is the Euler-Mascheroni constant.
}
\label{fig02}
\end{figure}

We next study the negativity between the impurity and the bath in the SIAM.
As the Anderson impurity has both spin and charge fluctuations,
the negativity can be affected by the both.

In \Fig{fig02} we show the negativity $\mc{N}$ between the whole degrees (spin and charge) of freedom of the impurity and the bath. The negativity $\mc{N}$ depends on $U$, reflecting the dependence of the SIAM on $U$.
The negativity $\mc{N}$ has a different value at zero temperature $T=0^+$.
Moreover,  $\mc{N}$ exhibits a crossover around $T=T_\mr{SC}$ for any value of $U$ and another crossover around $T=T_\mr{LM}$ for large U (e.g., $U=20\Gamma$).

At zero temperature $T=0^+$, the negativity $\mc{N}$ in~\Fig{fig02}(b) decreases with increasing $U$,
has a value $1$ for $U \rightarrow \infty$, and $3$ for $U=0$.
It happens since the charge fluctuation at the impurity is not completely suppressed (i.e., there is a finite probability that the impurity is empty or doubly occupied) for finite $U$ even at $T=0^+$.
One can understand the $U$-dependence of the negativity $\mc{N}(T = 0^+)$ in the two limits of $U \rightarrow \infty$ and $U = 0$ as follows.
In the limit of $U \to \infty$, the ground state of the SIAM is the Kondo singlet,
since the SIAM reduces to the SIKM at low temperature~\cite{Hewson97}.
Therefore, for $U \rightarrow \infty$, the SIAM has the same value $\mc{N} (T = 0^+) = 1$ as the SIKM.
In the limit of $U=0$, the SIAM is equivalent to two copies of the resonant level model of spinless fermions,
where each copy corresponds to the electron system of each spin.
Because of $\epsilon_\mr{d} = -U/2 = 0$, the ground state of each copy is a Bell state,
which is an equal-weight superposition of a state with the empty resonant level and the other state with the filled resonant level.
So the ground state of the SIAM at $U=0$ is a tensor product of two Bell states.
The negativity of this tensor product is $3$, which can be understood using the logarithmic negativity.
The logarithmic negativity $\log_2 (\mc{N}+1)$ is a monotone function of the negativity $\mc{N}$,
and the logarithmic negativity is additive though not convex~\cite{Plenio05}.
Each Bell state has the logarithmic negativity $\log_2 (\mc{N}+1) = \log_2 (1+1) = 1$.
Due to the additivity,
the logarithmic negativity is $2$ for the tensor product of the two Bell states.
$\log_2 (\mc{N}+1) = 2$ means that for $U=0$, the SIAM has the negativity $\mc{N}(T=0^+) = 3$.

At finite temperature $T$, the negativity $\mc{N}$ shows two kinks, one around $T=T_\mr{SC}$ and another around $T=T_\mr{LM}$ which indicate crossovers.
The crossover around $T = T_\mr{SC}$ occurs for any value of $U$,
while the crossover around $T = T_\mr{LM}$ appears only for sufficiently large $U$ (as for $U = 20\Gamma$).
In Fig.~\ref{fig02},
we show that the crossovers correspond to those of the impurity entropy $S_\mathrm{imp} \equiv S_\mathrm{tot}-S_\mathrm{bath}$, 
where $S_\mathrm{tot}$ ($S_\mathrm{bath}$) is the entropy 
of the impurity-bath system (of the bath only)~\cite{Bulla08}.
The plateaus in $S_\mathrm{imp}$ imply  the fixed points in the SIAM,
and the slanted lines connecting adjacent plateaus represent crossovers between the fixed points.
In the curve for $U=20\Gamma$ in \Fig{fig02}(c),
we observe three plateaus of  $S_\mathrm{imp}$
which have been interpreted as different fixed points:
The plateau at the highest $T$ means the free-orbital fixed point, where the charge degree of freedom of the impurity is not frozen and the spin degree of freedom of the impurity is weakly correlated to the bath.
The intermediate plateau indicates the local-moment fixed point where the charge degree of freedom becomes frozen (i.e., only the singly occupied impurity states involve in the fixed-point Hamiltonian) for large $U$ and the spin degree of freedom is still weakly correlated to the bath.
$S_\mr{imp}$ does not show clearly the intermediate plateau if $U/\Gamma$ is not sufficiently large
(e.g., when $U / \Gamma =10$ and $5$).
The plateau at the lowest $T$ corresponds to the strong-coupling fixed point in which
the spin degrees of freedom of the impurity is strongly entangled with the bath,
similarly to the strong-coupling fixed point in the SIKM.
In~\Fig{fig02}(c), $T=T_\mr{SC}$ is located at the end of the plateau for the strong-coupling fixed point for all values of $U$,
and $T = T_\mr{LM}$ is located at the end of the intermediate plateau (the local-moment fixed point) of the $S_\mr{imp}$ only for $U=20\Gamma$.
The comparison between $\mc{N}$ and $S_\mr{imp}$ shows that $\mc{N}$ captures the fixed points and the crossovers between them.

Note that the dependence of $\mc{N}(T = 0^+)$ vs.~$U$ is not contradictory to the interpretation of the local-moment and strong-coupling fixed points.
The impurity states away from single occupation are \textit{not} forbidden in these two fixed points;
they merely do not participate in the effective Hamiltonian of these fixed points.
Thus the NRG result of the ground state, which includes the empty and doubly occupied impurity states,
is consistent with the interpretation of the fixed points.

\begin{figure}
\centerline{\includegraphics[width=0.5\textwidth]{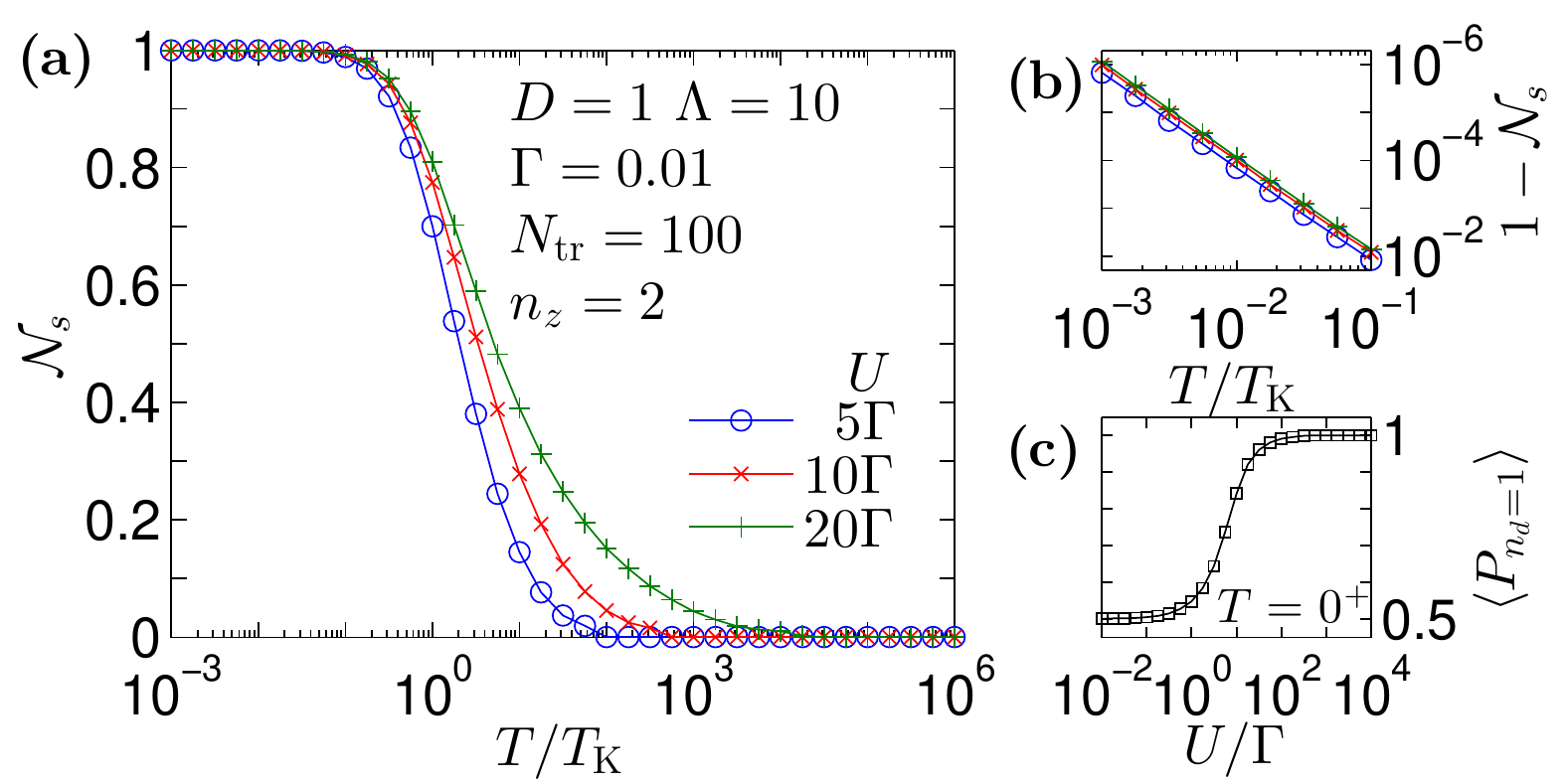}}
\caption{(Color online)
(a) Temperature ($T$) dependence of the negativity $\mc{N}_s$ quantifying the entanglement between the impurity \textit{spin} and the bath in the SIAM.
Contrary to $\mc{N}$ in \Fig{fig02}, $\mc{N}_s$ shows the same behavior as the negativity $\mc{N}$ of the SIKM shown in \Fig{fig01}(a).
At zero temperature $T = 0^+$, $\mc{N}_s$ is independent of $U$,
and around $T = T_\mr{LM}$, $\mc{N}_s$ does not exhibit any kink.
(b) At low temperature $T \ll T_\mr{K}$, $\mc{N}_s$ has a quadratic dependence on $T$,
similarly to the negativity $\mc{N}$ of the SIKM in \Fig{fig01}(b).
(c) The probability $\la P_{n_d = 1} \ra = \tr \rho_s$ that the impurity is singly occupied,
as a function of $U/\Gamma$. It increases as $U$ increases.
Here the Kondo temperature  $T_\mr{K}$ defined in \Fig{fig02} is used.
}
\label{fig03}
\end{figure}

Next we focus on the effect of the spin fluctuation on the entanglement between the impurity and the bath.
In \Fig{fig03} we compute the negativity $\mc{N}_s$ between the \textit{spin} degree of freedom of the impurity and the bath, after projecting out the doubly occupied and empty impurity states.
The negativity $\mc{N}_s$ shows the same behavior as the negativity $\mc{N}$ in the SIKM.
That is, $\mc{N}_s$ is defined as
\begin{align}
\mc{N}_s \equiv \mc{N}\big(\rho_s / \tr \rho_s \big),
\label{eq:Ns}
\end{align}
where $\rho_s \equiv P_{n_d = 1} \rho_T P_{n_d = 1}$, 
$\rho_T$ the thermal density matrix in \Eq{eq:rho_T}, 
and $P_{n_d = 1}$ the projector onto the subspace in which the impurity is half-filled, i.e.,
$n_d = \sum_\mu n_{d \mu} = 1$.
The doubly occupied and empty impurity states are projected out by applying the projector $P_{n_d = 1}$, 
so only the spin degree of freedom of the impurity remain.
Therefore, $\mc{N}_s = 1$ means that the impurity spin and the bath are maximally entangled,
 as in the SIKM case.

The negativity $\mc{N}_s(T=0^+)=1$ is independent of $U$,
which is due to the Kondo spin singlet formed by the impurity spin and the bath near the strong coupling fixed point.
At low temperature $T \ll T_\mr{K}$ near the strong-coupling fixed point, 
the negativity $\mc{N}_s$ in \Fig{fig03}(b) shows a universal quadratic scaling behavior $\mc{N}_s \simeq 1 - a_{\mc{N},\mr{1CK}} (T/T_\mathrm{K})^2$.
This scaling behavior is the same as that of the impurity-bath negativity $\mc{N}$ of the SIKM in~\Fig{fig01}(b).
Moreover,  $\mc{N}_s$ has no kink around $T=T_\mr{LM}$, since  the crossover around $T=T_\mr{LM}$, occuring between the local-moment fixed point and the free-orbital fixed point,
involves only the change in charge fluctuations.
 
It is natural that $\mc{N}_s$ in the SIAM shows the same behavior as $\mc{N}$ in the SIKM at low temperature, since the SIKM can be obtained from the SIAM by restricting the impurity to be half-filled or suppressing charge fluctuations.
In contrast, the impurity-bath negativity $\mc{N}$ of the SIAM does not show the low-temperature universal scaling 
because the charge fluctuation of the impurity does not participate in the universal Kondo physics.

In addition, we characterize the degree of the charge fluctuation at the impurity by using the probability $\la P_{n_d = 1} \ra = \tr \rho_s$ of the single occupancy at the impurity,
in \Fig{fig03}(c).
The single occupancy probability $\la P_{n_d = 1} \ra$ increases as $U$ increases,
since the charge fluctuation gets suppressed.
It is consistent with the $U$ dependence of the $\mc{N}(T=0^+)$ of the SIAM in~\Fig{fig02}(b).
In the limit $U \to \infty$,
the charge fluctuation is completely suppressed to compel the impurity to be half-filled, 
so $\mc{N}(T=0^+)=1$ and $\la P_{n_d=1} \ra = 1$.
In the opposite limit $U = 0$, the ground state is equivalent to the tensor product of two Bell states as discussed before.
In this case, $\la P_{n_d = 1} \ra = 1/2$, since the ground state can be represented as an equal superposition of the four state vectors whose impurity states are fully occupied, spin-up, spin-down, and empty, respectively.
 
\section{Error analysis}\label{600}

\begin{figure}
\centerline{\includegraphics[width=0.5\textwidth]{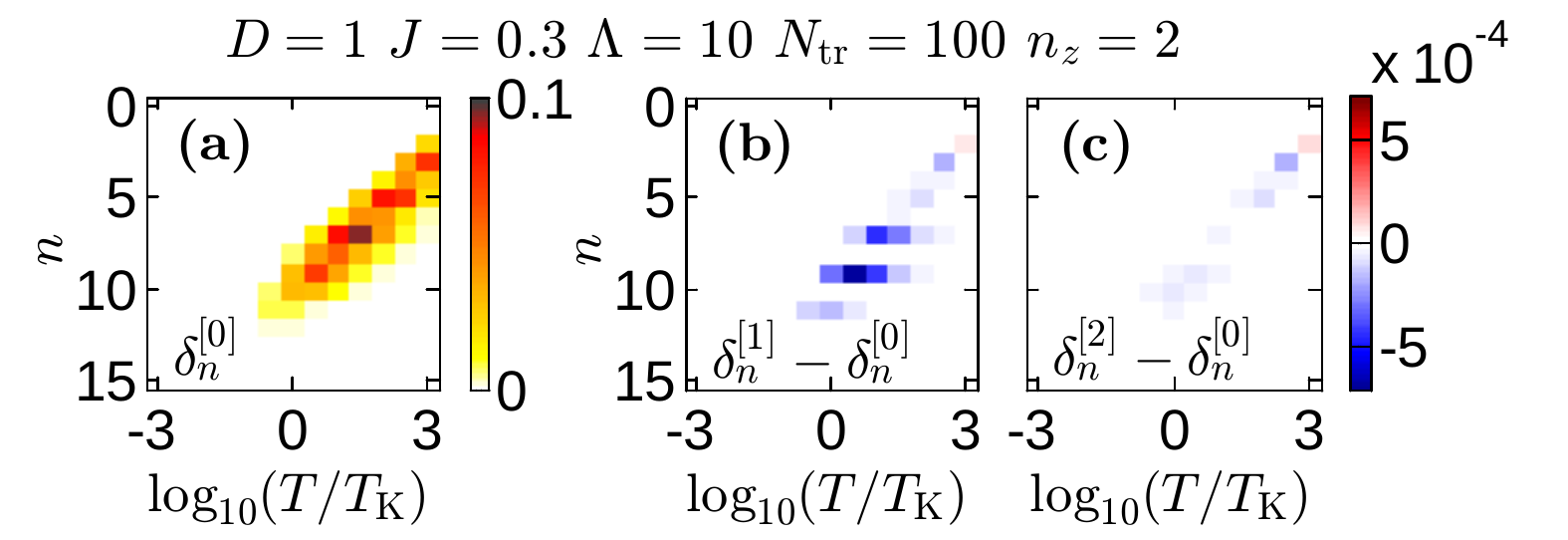}}
\caption{
(Color online) Estimation of errors in the NRG approximation for the SIKM.
(a) Plot of $\delta_n^{[0]}$ with varying $n$ and $T$.
At each $T$,
$\delta_n^{[0]}$ is the largest at $n \simeq -2 \log_\Lambda T$ as $\tr \rho_n^D$ is the largest thereat.
(b,c) Plot of $| \delta_n^{[k]} - \delta_n^{[0]} |$ with varying $n$ and $T$ for (b) $k=1$ and (c) $k=2$.
Both $| \delta_n^{[1]} - \delta_n^{[0]} |$ and $| \delta_n^{[2]} - \delta_n^{[0]} |$ are much smaller than $\delta_n^{[0]}$ by more than two orders of magnitude.
Note that $| \delta_n^{[2]} - \delta_n^{[0]} |$ is smaller than $| \delta_n^{[1]} - \delta_n^{[0]} |$,
which is a manifestation of an even-odd behaviour in the renormalization group flow, i.e., the finite-size energy spectrum.
The values at $n > 15$ are much smaller than those at $n < 15$, hence, they are not shown here.
}
\label{fig04}
\end{figure}

\begin{figure}
\centerline{\includegraphics[width=0.5\textwidth]{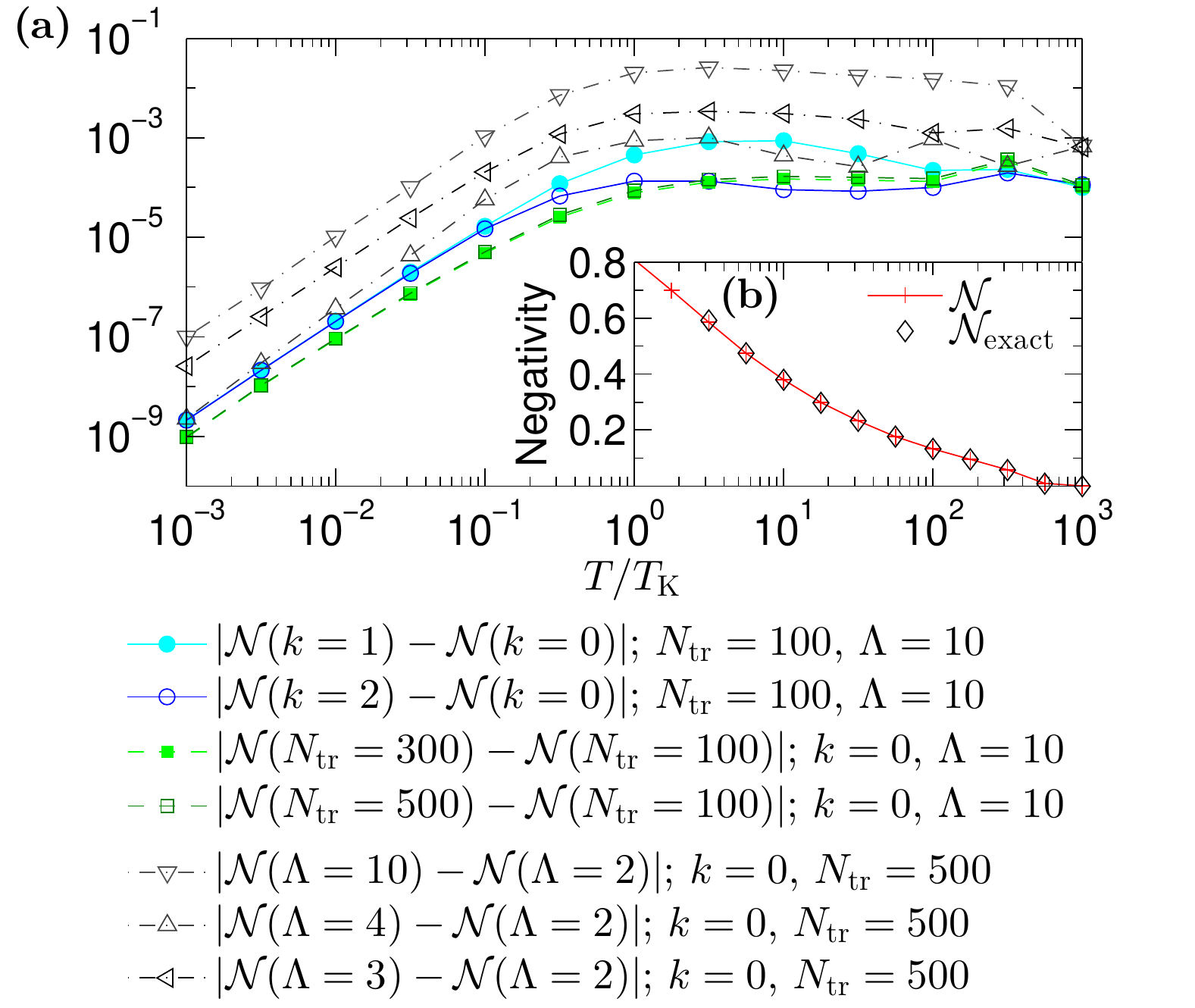}}
\caption{(Color online) 
Comparison of the computed values of $\mc{N} (T)$ from different numerical settings.
For consistency, we consider the SIKM with $J = 0.3$ and $n_z = 2$.
(a) The deviations of $\mc{N}(k, N_\mr{tr},\Lambda)$ for different parameters: the degree $k$ of the NRG approximation ($\delta_n \to \delta_n^{[k]}$),
truncation threshold $N_\mr{tr}$, or discretization parameter $\Lambda$.
The deviation $|\mc{N}(k=1,2) - \mc{N}(k=0)|$ is comparable or much smaller than the other deviations indicating NRG intrinsic errors, implying that the errors generated by the NRG approximation are negligible within the NRG intrinsic errors. The deviations indicating NRG intrinsic errors are maximal at $T \gtrsim T_\mr{K}$, but they are fairly small in comparison with $\mc{N}$.
(b) The negativity $\mc{N}$ computed via the NRG approximation, with choosing $k = 0$, $N_\mr{tr} = 100$, and $\Lambda = 10$, is compared with the exactly computed value $\mc{N}_\mr{exact}$ for $T > T_\mr{K}$.
Here $\mc{N}_\mr{exact}$ is obtained by exactly diagonalizing the Wilson chain consisting of the impurity and 7 bath sites.
$\mc{N}_\mr{exact}$ has only the discretization artifact due to the same $\Lambda = 10$. Note that the lowest energy scale of this short Wilson chain $\Lambda^{-6/2} = 10^{-3}$ is larger than the values of $T$ chosen for computing $\mc{N}_\mr{exact}$.
}
\label{fig05}
\end{figure} 

We analyze the errors in the negativity calculation subject to the NRG method.
For the SIKM, for example,
we  investigate how the computed value of $\mc{N}$ depends on
the NRG approximation, the truncation in the iterative diagonalization,
and the logarithmic discretization.

We first estimate how the NRG approximation affects the value of $\mc{N}$.
Under the NRG approximation in Eq.~\eqref{eq:deltB00},
we replace $\mc{R}_n$ and $\delta_n$ by $\rho_n^D$
and $\delta_n^{[0]}$, respectively, where the information of the chain site $n' > n$ is traced out.
This approximation can be improved by replacing $\mc{R}_n$ and $\delta_n$ by $\rho_n^D$
and $\delta_n^{[k]}$, respectively, where  the information of the chain site $n' > n +k $ is traced out.
The expression of $\delta_n^{[k]}$ is
\begin{align}
{\delta}_n^{[k]} &\equiv \mc{N} \bigg( \tr_{n+k+1, \cdots, N} \Big[ \mc{R}_n \Big] \bigg) \nonumber \\
& \quad + \mc{N} \bigg( \tr_{n+k+1, \cdots, N} \Big[ \sum_{n'>n}^N \mc{R}_{n'} \Big] \bigg) \nonumber \\
& \quad - \mc{N} \bigg( \tr_{n+k+1, \cdots, N} \Big[ \mc{R}_n + \sum_{n'>n}^N \mc{R}_{n'} \Big] \bigg) \nonumber \\
&= \mc{N} \Big( \rho_n^D \otimes I_{n+1} \otimes \cdots \otimes I_{n+k} \Big) \nonumber \\
& \quad + \mc{N} \Big( \sum_{n'>n}^{n+k} \rho_{n'}^D \otimes I_{n'+1} \otimes \cdots \otimes I_{n+k} + \rho_{n+k}^K \Big)
\nonumber \\
&\quad - \mc{N} \Big( \sum_{n'=n}^{n+k} \rho_{n'}^D \otimes I_{n'+1} \otimes \cdots \otimes I_{n+k} + \rho_{n+k}^K \Big), \label{eq:deltak}
\end{align}
where $k = 0, 1, 2, \cdots$.
For $k = 0$, \Eq{eq:deltak} reduces to \Eq{eq:deltB00}.
For larger $k$, less information is traced out so that $\mc{N}$ can be computed more precisely,
however, the computation cost rapidly increases; as $k \to \infty$, the calculation becomes exact within the NRG method.
Note that the replacement of $\mc{R}_n$ by $\rho_n^D$ is not affected although less information is traced out, because
\begin{align}
\mc{N}(\mc{R}_n) = \mc{N}(\tr_{n+k+1, \cdots, N} [ \mc{R}_n ] ) = \mc{N}(\rho_n^D) .
\end{align}
In \Fig{fig04}, we show the magnitudes of $\delta_n^{[0]}$ and of the deviations $\delta_n^{[k]} - \delta_n^{[0]}$ for $k = 1, 2$.
In \Fig{fig05}, we display $|\mc{N} (k) - \mc{N} (k = 0)|$ for $k=1,2$, where  $\mc{N}(k)$ is the computation of $\mc{N}$ with the approximation of replacing $\delta_n$ by $\delta_n^{[k]}$.
$|\mc{N} (k =1,2) - \mc{N} (k = 0)|$ is at most $O(10^{-3})$ for $T \gtrsim T_\mr{K}$,
and scale as $\sim 10^{-3} \times (T / T_\mr{K})^2$ for $T \ll T_\mr{K}$, showing that $|\mc{N} (k =1,2) - \mc{N} (k = 0)|$ is negligibly small.
These verify that the NRG approximation of $\delta_n \to \delta_n^{[0]}$ is already good enough.

We next check the change of $\mc{N}$ with varying an NRG parameter $N_\textrm{tr}$,
the number of the kept states in each iteration step.
As shown in \Fig{fig05},
the change is negligible, showing that $\mc{N}$ is almost independent of $N_\textrm{tr}$.
We notice that
the change is comparable with $|\mc{N} (k =1,2) - \mc{N} (k = 0)|$.
This is natural, since both of choosing smaller $N_\textrm{tr}$ and smaller $k$ lead to common errors due to neglecting the information of a later part of the NRG chain.
This observation suggests that the amount of errors in computing $\mc{N}$ due to the NRG approximation can be estimated by the change $\mc{N}$ with varying $N_\textrm{tr}$.
This will provide a practical approach to estimate the errors due to the NRG approximation in general systems such as the multi-channel Kondo model,
where the direct calculations of $\delta_n^{[k]}$ ($k > 0$) are hardly feasible.

We also check the change of $\mc{N}$ with varying the NRG discretization parameter $\Lambda$.
The change is also negligible in comparison with $\mc{N}$.
Note that the change of $\mc{N}$ with $\Lambda$ is larger than 
that with $N_\textrm{tr}$ and $k$.
It is because different values of $\Lambda$ yield different discretized Hamiltonians.

The accuracy of our computation of $\mc{N}$  can be also tested  at $T > T_\textrm{K}$.
In this temperature range, the relevant length (less than 7) of the Wilson chain is so short that $\mc{N}$ can be computed exactly by diagonalizing the whole NRG chain.
Figure~\ref{fig05}(b) shows that our computation of $\mc{N}$ with the NRG approximation is almost identical to the values obtained by the exact diagonalization.

All the above observations demonstrate that our computation of $\mc{N}$ with the NRG approximation is sufficiently accurate.

\section{Conclusion}\label{700}

We develop the NRG method for computing the negativity $\mc{N}$ quantifying an impurity-bath entanglement in a quantum impurity system at finite temperature, and apply it to the SIKM and the SIAM.
For the SIKM, the $T$-dependence of $\mc{N}$ shows the universal power-law scaling at low temperature,
and the sudden death at high temperature.
For the SIAM, $\mc{N}$ is affected by both the spin and charge fluctuations at the impurity.
The spin fluctuation causes $\mc{N}$ to show a universal power-law scaling behavior similar to the SIKM.
The negativity $\mc{N}$ depends on $U$ even at zero temperature,
indicating that the charge fluctuation survives even near the strong-coupling fixed point for finite $U$.

Since the error due to the NRG approximation is smaller than the other artifacts intrinsic to the NRG, 
our computation of $\mc{N}$ is sufficiently accurate.
In this sense, the current scheme for computing the negativity is advantageous over the earlier one for the EoF~\cite{Lee15}:
The latter could only provide the lower and upper bounds of entanglement,
and the interval between these bounds can exceed the intrinsic errors in the NRG.
We anticipate that our method will be applicable to general quantum impurity systems in various situations and reveal entanglement perspective in understanding them.

\begin{acknowledgments}
We thank A. Weichselbaum for fruitful discussion.
H.-S.S. and J.S. are supported by Korea NRF (Grant Nos. 2015R1A2A1A15051869 and 2016R1A5A1008184).
S.-S.B.L. acknowledges support from the Alexander von Humboldt Foundation and the Carl Friedrich von Siemens Foundation.
\end{acknowledgments}

\appendix

\section{Scaling behavior at low tmperature}\label{A00}
We derive the scaling behavior of the impurity-bath negativity in \Eq{eq:Nscaling} for the SIKM at low $T\ll T_{\mathrm{K}}$ using the bosonization. 
This scaling behavior originates from the low-energy excitations of the Fermi-liquid quasiparticles in the SIKM.

We set the thermal density matrix $\rho = \sum_i w_i |E_i\rangle\langle E_i|$ in terms of the energy eigenstate $|E_i\rangle$ of the SIKM with energy $E_i$ and the Boltzmann factor $w_i$ of $|E_i\rangle$ satisfying $\sum_i w_i = 1$.
$\rho$ can be approximated by the eigenstates $\{ |E_i \rangle \}$ satisfying $E_i \sim T$, because $w_i$ decreases exponentially in $E_i / T$ while state degeneracy increases algebraically in $E_i$.
 
 To compute $\mc{N}$, we represent $\rho$ in a bipartite basis of $\{|\mu\rangle \otimes |\phi_{i\eta}\rangle\}$, where $\{ |\mu\rangle \}$ ($\{ |\phi_{i\eta}\rangle \}$) is the orthonormal impurity (bath) basis.
Using the bosonization~\cite{Zarand00} and the effective theory near the strong-coupling fixed point~\cite{Hewson97},
we represent the eigenstate $|E_i\rangle$ 
as~\cite{Lee15}
\begin{align}
|E_i \rangle = \frac{1}{\sqrt{2}}\sum_{\mu = \uparrow, \downarrow}|\mu\rangle \otimes ( |\phi_{i\mu}\rangle + |\chi_{i\mu}\rangle ) ,
\label{eq:Eb}
\end{align}
where $\langle E_i | E_{i'}\rangle = \delta_{ii'}$ and $\langle \phi_{i\eta}| \phi_{i'\eta'}\rangle = \delta_{ii'}\delta_{\eta\eta'}$.
$\{|\chi_{i\eta}\rangle\}$ are bath states of $|\chi_{i\eta}\rangle \in \mr{span}\{ | \phi_{i\eta} \rangle \}$, satisfying $\langle \chi_{i\eta}| \phi_{i\eta}\rangle = 0$, and $\sqrt{ \langle \chi_{i\eta}| \chi_{i'\eta'}\rangle } \sim \langle\chi_{i\eta}|\phi_{i'\eta'}\rangle \sim O(T/T_\mr{K})$.
The latter relation is due to the Fermi-liquid behavior of the SIKM at low $T$, and
it determines the scaling exponent of the negativity.
Applying \Eq{eq:Eb}, we write the density matrix $\rho$ as
\begin{align}
\rho = \sum_{ii'}\sum_{\mu,\mu',\eta,\eta' = \uparrow,\downarrow} [\rho]_{(\mu, i, \eta), (\mu', i',\eta')} |\mu\rangle\langle\mu'| \otimes |\phi_{i\eta}\rangle\langle \phi_{i'\eta'}| ,
\label{eq:rhob}
\end{align}
whose element is
\begin{align}\nonumber
[\rho]_{(\mu, i, \eta), (\mu', i',\eta')} &= \sum_{j}\frac{w_j}{2} 
\big[ \delta_{ij}\delta_{\eta\mu} + \langle \phi_{i\eta} | \chi_{j\mu}\rangle \big]
\\
&\qquad \times \big[ \delta_{ji'}\delta_{\mu'\eta'} + \langle \chi_{j\mu'}| \phi_{i'\eta'}\rangle \big] .
\end{align}

To obtain the negativity using \Eq{eq:Nrho}, we need to compute $\mr{Tr}|\rho^{T_A}|$, where $\rho^{T_A}$ is
\begin{align}
\rho^{T_A} = \sum_{ii'}\sum_{\mu,\mu',\eta,\eta' = \uparrow,\downarrow} [\rho]_{(\mu, i, \eta), (\mu', i',\eta')} |\mu'\rangle\langle\mu| \otimes |\phi_{i\eta}\rangle\langle \phi_{i'\eta'}|.
\label{eq:rhoTb}
\end{align}
$\mr{Tr}|\rho^{T_A}|$, the sum of the singular values $\sigma_{\mu i \eta}$ of $\rho^{T_A}$, 
equals the sum of the square root of the singular values $\sigma_{\mu i \eta}^2$ of $(\rho^{T_A})^2$.
We compute the singular values of $(\rho^{T_A})^2$, since they are easier to be estimated.
Using the facts that (i) the leading order and the next leading order of the diagonal terms of $(\rho^{T_A})^2$ are $O(1)$ and $O(T^2 / T^2_\textrm{K})$, respectively, (ii) the leading order of the off-diagonal terms of $(\rho^{T_A})^2$ are $O(T / T_\textrm{K})$, and (iii) $T / T_\textrm{K} \ll 1$, we compute the singular values $\sigma_{\mu i \eta}^2$ of $(\rho^{T_A})^2$ and find
\begin{align}
\sigma_{\mu i \eta} = c_{\mu i \eta} + c'_{\mu i \eta} (T/T_\mathrm{K})^2 + \cdots,
\end{align}
where $c_{\mu i \eta}$ and $c'_{\mu i \eta}$ are coefficients of order O(1).
Then, the impurity-bath negativity $\mc{N}(\rho)$ is obtained as
\begin{align}
\mc{N}(\rho) &= \mathrm{Tr}|\rho^{T_A}| - \tr\, \rho = \sum_{\mu i \eta} \sigma_{\mu i\eta} - 1 \nonumber \\
& = c + a' \left(T/T_\mathrm{K}\right)^2,
\end{align}
where $c$ and $a'$ are constants.
Using the property of the SIKM that $\mc{N}=1$ at $T=0$ and it cannot increase with increasing $T$,
we obtain Eq.~\eqref{eq:Nscaling} at low $T \ll T_{\mathrm{K}}$,
\begin{align}
\mathcal{N} \simeq 1 - a_{\mc{N},\mr{1CK}} \left(T/T_\mathrm{K}\right)^2 ,
\end{align}
where a coefficient $a_{\mc{N},\mr{1CK}} >0$ is $O(1)$.

\section{Sudden death in the Impurity-Bath Entanglement}\label{B00}

Here we explain the linear dependence of the sudden death temperature $T_\mr{SD} \sim J$ in the SIKM result of \Fig{fig01}(c),
by considering the Wilson chain with only one bath site, i.e., $N = 0$, as a minimal model.
For this minimal model, both the negativity and the EoF yields the same sudden death temperature $T_\mr{SD} = J / \ln 3$.
Note that there is no bound entanglement at $T_\mr{SD}$, as the EoF, which can detect any bound entanglement, vanishes at $T_\mr{SD}$.

The energy eigenvalues and eigenstates of the Hamiltonian $H_{N=0}^\mr{SIKM}$ are given by:
\begin{equation}
\begin{tabular}{| c | c |}
\hline
\,\,\, Eigenvalue \,\,\, & Eigenstate \\
\hline
$-3J/4$ & \,\,\, $(\ket{{\Uparrow}}\ket{\da} - \ket{{\Downarrow}}\ket{\ua})/\sqrt{2}$ \,\,\, \\
\hline
\multirow{3}{*}{$J/4$} & $\ket{{\Uparrow}}\ket{\ua}$ \\
& $\ket{{\Downarrow}}\ket{\da}$ \\
& $(\ket{{\Uparrow}}\ket{\da} + \ket{{\Downarrow}}\ket{\ua})/\sqrt{2}$ \\
\hline
\multirow{4}{*}{$0$} & $\ket{{\Uparrow}}\ket{\ua\da}$ \\
& $\ket{{\Uparrow}}\ket{0}$ \\
& $\ket{{\Downarrow}}\ket{\ua\da}$ \\
& $\ket{{\Downarrow}}\ket{0}$ \\
\hline
\end{tabular}
\label{eq:eigen}
\end{equation}
Here $\ket{{\Uparrow}}$ and $\ket{{\Downarrow}}$ are the impurity spin state, and $\ket{0}$, $\ket{\ua}$, $\ket{\da}$, and $\ket{\ua\da}$ indicate the empty, spin-up, spin-down, and doubly occupied states of the electron bath site, respectively.
Then we construct the thermal density matrix $\rho_{0}^\mr{SIKM} = e^{-H_{N=0}^\mr{SIKM}/T} / \tr \, e^{-H_{N=0}^\mr{SIKM}/T}$ based on the eigendecomposition above.

First, for the negativity, one can directly apply \Eq{eq:Nrho} to the $\rho^\mr{SIKM}_0$ to obtain
\begin{align}
\mathcal{N}(\rho^\mathrm{SIKM}_0) = \mathrm{max}\Big(\frac{1 - 3e^{-J/T}}{1 + 4e^{-3J/4T} + 3e^{-J/T}}, \, 0 \Big) .
\end{align}
The negativity $\mc{N}(\rho^\mr{SIKM}_0)$ suffers sudden death at $T_\mr{SD} = J/\ln 3$.

On the other hand, the EoF is defined as an optimization problem,
\begin{equation}
\mc{E}_\mr{F}(\rho) \equiv \inf_{\{p_i, \ket{\psi_i}\}} \sum_i p_i \, \mc{E}_\mr{E} (\ket{\psi_i}),
\label{eq:EoF}
\end{equation}
where $\mc{E}_\mr{E} (\ket{\psi_i}) = - \tr \rho_{iA} \log_2 \rho_{iA}$ is the entanglement entropy of $\ket{\psi_i}$, and
$\rho_{iA} = \tr_B \prj{\psi_i}$ is the reduced density matrix in which the bath $B$ is traced out.
That is, the EoF for a mixed state $\rho$ is the infimum of the weighted sum of the entanglement entropy, $\sum_i p_i \mc{E}_\mr{E} (\ket{\psi_i})$,
over all possible pure-state decomposition $\rho = \sum_i p_i \prj{\psi_i}$.
Here $\ket{\psi_i}$'s are normalized, i.e., $\ovl{\psi_i}{\psi_i} = 1$, but do not need to be orthogonal to each other.
As mentioned in \Sec{100}, there is no general solution of \Eq{eq:EoF}.
But fortunately for $\rho_0^\mr{SIKM}$, there exists an analytic solution,
which we will derive by the following steps.

(i) The density matrix $\rho_0^\mr{SIKM}$ can be decomposed into a block diagonal form,
\begin{equation}
\rho_0^\mr{SIKM} = \rho_1 + \rho_2,
\label{eq:rho_block}
\end{equation}
where $\rho_1 \in \mc{H}_1 \equiv \mr{span}\{\ket{{\Uparrow}}, \ket{{\Downarrow}}\} \otimes \mr{span}\{\ket{{\uparrow}}, \ket{{\downarrow}}\}$
and $\rho_2 \in \mc{H}_2 \equiv \mr{span}\{\ket{{\Uparrow}}, \ket{{\Downarrow}}\} \otimes \mr{span}\{\ket{0}, \ket{\ua\da}\}$.
The bath site is half filled in the subspace $\mc{H}_1$,
while empty or doubly occupied in $\mc{H}_2$.
In other words, $\mc{H}_2$ is spanned by the energy eigenstates with zero eigenvalues, and $\mc{H}_1$ by the rest.

(ii) Consider a pure state
\begin{equation}
\ket{\varphi} = c_1 \ket{\varphi_1} + c_2 \ket{\varphi_2}
\end{equation}
for arbitrary normalized states $\ket{\varphi_1} \in \mc{H}_1$ and $\ket{\varphi_2} \in \mc{H}_2$,
where $c_1$ and $c_2$ are complex numbers satisfying $|c_1|^2 + |c_2|^2 = 1$.
Since the bath states of $\ket{\varphi_1}$ and $\ket{\varphi_2}$ are orthogonal by construction,
we have 
\begin{equation}
\tr_B \prj{\varphi} = |c_1|^2 \tr_B \prj{\varphi_1} + |c_2|^2 \tr_B \prj{\varphi_2} .
\end{equation}
Then the concavity of the von Neumann entropy leads to an inequality
\begin{equation}
\mc{E}_E(\ket{\varphi}) \geq |c_1|^2 \mc{E}_E(\ket{\varphi_1}) + |c_2|^2 \mc{E}_E(\ket{\varphi_2}) .
\end{equation}
Based on the block diagonal form in \Eq{eq:rho_block} and this concavity,
we find a restriction to the optimal pure-state decomposition $\rho_0^\mr{SIKM} = \sum_i p_i^\mr{op} \prj{\psi_i^\mr{op}}$,
which provides $\mc{E}_\mr{F} (\rho_0^\mr{SIKM}) = \sum_i p_i^\mr{op} \mc{E}_\mr{E} (\ket{\psi_i^\mr{op}})$:
Each state $\ket{\psi_i^\mr{op}}$ should be in either $\mc{H}_1$ or $\mc{H}_2$, not a superposition of a state in $\mc{H}_1$ and another in $\mc{H}_2$.
(It can be proven by contradiction.)
Therefore, the EoF reduces to
\begin{equation}
\begin{aligned}
\mc{E}_\mr{F} (\rho_0^\mr{SIKM}) &= \mc{E}_\mr{F} (\rho_1) + \mc{E}_\mr{F} (\rho_2) \\
&= \mc{E}_\mr{F} (\rho_1) \\
&= \tr \, \rho_1 \cdot \mc{E}_\mr{F} (\rho_1 / \tr \, \rho_1)
\end{aligned}
\end{equation}
where at the second equality we used $\mc{E}_\mr{F} (\rho_2) = 0$ since $\rho_2$ is the mixture of product states [see \Eq{eq:eigen}],
and at the last equality we pulled out the normalization factor
\begin{equation}
\tr \, \rho_1 = \frac{ e^{3J/4T} + 3 e^{J/4T} }{ e^{3J/4T} + 3 e^{J/4T} + 4} ,
\end{equation}
for convenience below.

(iii) We can regard $\rho_1$ as the state of two qubits;
now we can use the concurrence~\cite{Wootters98} to derive the EoF of the normalized state $\rho_1 / \tr \, \rho_1$,
\begin{gather}
\mc{E}_\mr{F} \left( \frac{\rho_1}{\tr \, \rho_1} \right) = h \left( \frac{1 + \sqrt{1 - \mc{C}^2}}{2} \right), \label{eq:EoF_conc}
\end{gather}
where $h(x) = -x \log_2 x - (1-x) \log_2 (1-x)$ and $\mc{C}$ is the concurrence of $\rho_1 / \tr \, \rho_1$.
Here the right-hand side expression of \Eq{eq:EoF_conc} is a monotonically increasing function of $\mc{C}$.
The concurrence is given by
\begin{equation}
\mc{C} = \mr{max} \Big( \frac{e^{J/T} - 3}{e^{J/T} + 3}, \, 0 \Big)
\end{equation}
which indicates that $\mc{E}_\mr{F} (\rho_1 / \tr \, \rho_1)$, and $\mc{E}_\mr{F}(\rho_0^\mr{SIKM})$ also, suffer the sudden death at $T_\mr{SD} = J / \log 3$.
Both the negativity and the EoF yield the same $T_\mr{SD}$,
which means that there is no bound entanglement.
It is natural, since  the entanglement of $\rho_0^\mr{SIKM}$ is contributed only from $\rho_1$ that can be regarded as a two-qubit state,
and there is no bound entanglment for two qubits in general.

\bibliography{EN_NRG_ref}

\end{document}